%% file: main.tex
\documentclass[acmsmall,screen,submit]{acmart}
\usepackage{tabularx}
\usepackage{hyperref}
\usepackage{cleveref}
\usepackage{diagbox}
\usepackage{multirow}
\usepackage{caption}
\usepackage{graphicx}
\usepackage{float} 
\usepackage{subcaption}
\usepackage{makecell}
\usepackage[most]{tcolorbox}
\PassOptionsToPackage{prologue,dvipsnames}{xcolor}

\newcommand{\rqbox}[1]{

\begin{tcolorbox}[tile, size=fbox, boxsep=2mm, boxrule=0pt, top=0pt, bottom=0pt,
borderline west={1mm}{0pt}{blue!50!white}, colback=blue!5!white]
#1
\end{tcolorbox}

}
\usepackage{xcolor} %

\newcommand{\lky}[1]{\textcolor{black}{#1}}

\newcommand{\tablecolor}{black}

\crefname{figure}{Fig.}{Fig}
\crefname{table}{Table}{Table}
\crefname{equation}{Eq.}{eq}
\crefname{section}{Section}{Section}

\newcolumntype{L}[1]{>{\raggedright\arraybackslash}p{#1}}
 
\newcolumntype{C}[1]{>{\centering\arraybackslash}p{#1}}
 
\newcolumntype{R}[1]{>{\raggedleft\arraybackslash}p{#1}}
\AtBeginDocument{%
  }

\setcopyright{acmlicensed}
\acmDOI{10.1145/3729357}
\acmYear{2025}
\acmJournal{PACMSE}
\acmVolume{2}
\acmNumber{FSE}
\acmArticle{FSE087}
\acmMonth{7}
\received{2025-02-25}
\received[accepted]{2025-04-01}

\begin{document}

\title{Zero-Shot Cross-Domain Code Search without Fine-Tuning}

\author{Keyu Liang}
\orcid{0009-0000-4613-247X}
\affiliation{%
  \department{The State Key Laboratory of Blockchain and Data Security}
  \institution{Zhejiang University}
  \city{Hangzhou}
  \country{China}
}
\email{liangkeyu@zju.edu.cn}

\author{Zhongxin Liu}
\orcid{0000-0002-1981-1626}
\authornote{Zhongxin Liu is the corresponding author and is also with Hangzhou High-Tech Zone (Binjiang) Institute of Blockchain and Data Security}
\affiliation{%
  \department{The State Key Laboratory of Blockchain and Data Security}
  \institution{Zhejiang University}
  \city{Hangzhou}
  \country{China}
}
\email{liu_zx@zju.edu.cn}

\author{Chao Liu}
\orcid{0000-0002-8283-9146}
\affiliation{%
  \department{School of Big Data and Software Engineering}
  \institution{Chongqing University}
  \city{Chongqing}
  \country{China}
}
\email{liu.chao@cqu.edu.cn}

\author{Zhiyuan Wan}
\orcid{0000-0001-7657-6653}
\affiliation{%
  \department{The State Key Laboratory of Blockchain and Data Security}
  \institution{Zhejiang University}
  \city{Hangzhou}
  \country{China}
}
\email{wanzhiyuan@zju.edu.cn}

\author{David Lo}
\orcid{0000-0002-4367-7201}
\affiliation{%
  \department{School of Computing and Information Systems}
  \institution{Singapore Management University}
  \city{Singapore}
  \country{Singapore}
}
\email{davidlo@smu.edu.sg}

\author{Xiaohu Yang}
\orcid{0000-0003-4111-4189}
\affiliation{%
  \department{The State Key Laboratory of Blockchain and Data Security}
  \institution{Zhejiang University}
  \city{Hangzhou}
  \country{China}
}
\email{yangxh@zju.edu.cn}

\begin{abstract}
Code search is a crucial task in software engineering, aiming to retrieve code snippets that are semantically relevant to a natural language query. Recently, Pre-trained Language Models (PLMs) have shown remarkable success and are widely adopted for code search tasks. However, PLM-based methods often struggle in cross-domain scenarios. When applied to a new domain, they typically require extensive fine-tuning with substantial data. Even worse, the data scarcity problem in new domains often forces these methods to operate in a zero-shot setting, resulting in a significant decline in performance. RAPID, which generates synthetic data for model fine-tuning, is currently the only effective method for zero-shot cross-domain code search. Despite its effectiveness, RAPID demands substantial computational resources for fine-tuning and needs to maintain specialized models for each domain, underscoring the need for a zero-shot, fine-tuning-free approach for cross-domain code search.

The key to tackling zero-shot cross-domain code search lies in bridging the gaps among domains. In this work, we propose to break the query-code matching process of code search into two simpler tasks: query-comment matching and code-code matching. We first conduct an empirical study to investigate the effectiveness of these two matching schemas in zero-shot cross-domain code search. Our findings highlight the strong complementarity among the three matching schemas, i.e., query-code, query-comment, and code-code matching. Based on the findings, we propose CodeBridge, a zero-shot, fine-tuning-free approach for cross-domain code search. Specifically, CodeBridge first employs zero-shot prompting to guide Large Language Models (LLMs) to generate a comment for each code snippet in the codebase and produce a code for each query. Subsequently, it encodes queries, code snippets, comments, and the generated code using PLMs and assesses similarities through three matching schemas: query-code, query-comment, and generated code-code. Lastly, CodeBridge leverages a sampling-based fusion approach that combines these three similarity scores to rank the final search outcomes. Experimental results show that our approach outperforms the state-of-the-art PLM-based code search approaches,  i.e., CoCoSoDa and UniXcoder, by an average of 21.4\% and 24.9\% in MRR, respectively, across three datasets. Our approach also yields results that are better than or comparable to those of the zero-shot cross-domain code search approach RAPID, which requires costly fine-tuning.
\end{abstract}



\received{1 June 2025}
\received[revised]{1 June 2025}
\received[accepted]{1 June 2025}

\maketitle

\input{sections/introduction} %
\input{sections/related_work}
\input{sections/empirical_setup}
\input{sections/empirical_result}
\input{sections/approach}

\input{sections/rq}

\input{sections/rq_result}
\input{sections/discussion}

\input{sections/threats}

\input{sections/conclusion}
\input{sections/data_avaliability}
\begin{acks}

This research/project is supported by Zhejiang Provincial Natural Science Foundation of China (No. LZ25F020003), the National Natural Science Foundation of China (No. 62202420, No. 62202074), and the National Research Foundation, under its Investigatorship Grant (NRF-NRFI08-2022-0002). Any opinions, findings and conclusions or recommendations expressed in this material are those of the author(s) and do not reflect the views of National Research Foundation, Singapore.

\end{acks}

\bibliographystyle{ACM-Reference-Format}
\bibliography{acmart}

\appendix

\end{document}

%% file: sections/introduction.tex
\section{Introduction}
Code search aims to retrieve the code snippets that are semantically relevant to a provided natural language query from a codebase.
It is one of the most frequent activities in software development~\cite{xia2017developers}, and can greatly enhance the efficiency of developers by assisting them in reusing the code from existing code repositories~\cite{liu2021opportunities,sachdev2018retrieval}.
\lky{In the era of large language models (LLMs), code search has gained new significance as a key component in retrieval-augmented generation (RAG)~\cite{lewis2020retrieval} and in-context learning~\cite{brown2020language} which are widely used to improve tasks like code generation~\cite{zhang2023repocoder,patel2024evaluating}.}
To this end, many approaches have been proposed to improve the effectiveness and efficiency of code search~\cite{lv2015codehow,gu2018deep,feng2020codebert,guo2022unixcoder}.

Early studies leverage information retrieval techniques for code search~\cite{lv2015codehow,bajracharya2014sourcerer,liu2021codematcher}. 
These methods primarily employ unsupervised text-matching algorithms, e.g., BM25~\cite{robertson2009probabilistic}, to match queries and code snippets.
However, these methods have proved insufficient for capturing deep semantics of queries and code snippets, limiting their effectiveness~\cite{gu2018deep}.
To better understand such semantics, prior works propose building neural models to encode queries and code snippets as embeddings and match queries and code snippets based on the similarity between their embeddings~\cite{gu2018deep,shuai2020improving,cheng2022csrs}.
Recently, due to the impressive understanding ability of pre-trained language models (PLMs)~\cite{feng2020codebert,guo2020graphcodebert,wang2021codet5,guo2022unixcoder,wang2023codet5+}, researchers propose fine-tuning pre-trained language models with high-quality query-code pairs for code search~\cite{li2022coderetriever,shi2023cocosoda} and achieved state-of-the-art performance.
We refer to such methods as PLM-based methods.

However, recent research shows that PLM-based methods exhibit significant performance degradation when applied to a domain, e.g., a new programming language, where they haven't been fine-tuned~\cite{fan2024rapid}.
A straightforward remedy is to collect query-code pairs from the new domain and fine-tune the pre-trained model with these data.
Specifically, query-code pairs can be either synthesized with comment-code pairs collected from code repositories or constructed manually.
However, in practice, the synthesis of high-quality query-code pairs often suffers from the shortage of code comments in software projects~\cite{spinellis2010code,briand2003software} and the prevalence of documentation issues~\cite{aghajani2020software,steidl2013quality}.
Such data scarcity can be more severe for low-resource domains.
On the other hand, fine-tuning PLMs requires a sufficient number of query-code pairs, and thus it is costly to construct them manually.
These constraints lead to a more practical code search scenario, where we perform cross-domain code search without using any query-code pair from the target domain.
We refer to this scenario as zero-shot cross-domain code search.

Due to the gap between domains, such as the distinct characteristics of various programming languages, it is challenging to develop a universal approach for zero-shot cross-domain code search.
Currently, RAPID \cite{fan2024rapid} is the only effective method capable of handling zero-shot cross-domain code search.
It involves generating pseudo queries for code snippets from the target domain using generative models and subsequently fine-tuning PLMs as retrieval models with the synthesized data.
While this method yields satisfactory results, it presents several limitations.
First, the fine-tuning process demands significant computational resources~\cite{shi2023towards}.
Second, each unique domain requires a specialized model, leading to increased financial costs and management complexity.
Considering the aforementioned issues, a \textbf{zero-shot, fine-tuning-free approach for cross-domain code search} would be more practical and appealing.

The challenge of zero-shot cross-domain code search lies in the domain gaps, which make the query-code mapping knowledge that code search models learned in some domains cannot be directly adopted to another domain.
Thus, the key to tackling this challenging task is bridging the domain gaps.
To bridge the gap, our idea is that the query-code matching process of code search, which involves both natural language understanding and programming language comprehension \cite{liu2021opportunities}, can be broken down into two easier tasks: query-comment matching and code-code matching.
Query-comment matching refers to retrieving code comments for a given query and returning the associated code, while code-code matching denotes retrieving the target code snippet when given another code snippet example that satisfies the query. 
These two matching schemas operate at similar levels of abstraction, which may help reduce the reliance on domain-specific mapping knowledge, thereby potentially mitigating the impact of domain gaps.

To explore the feasibility of this idea and understand why it can work, we first conduct an empirical study in a zero-shot cross-domain scenario.
We mainly focus on three key research questions (RQs):
\begin{itemize}
    \item RQ1: How effective is query-comment matching compared to query-code matching?
    \item RQ2: How effective is code-code matching compared to query-code matching?
    \item RQ3: Can the three matching schemas complement each other?
\end{itemize}
Through the empirical study, we find that:
\textbf{(1) Query-comment matching and code-code matching outperform query-code matching in certain cases.
(2) There is a high degree of complementarity among the three matching schemas, i.e., query-code, query-comment, and code-code. }

Inspired by these findings, we propose \textbf{CodeBridge}, a zero-shot, fine-tuning-free approach for cross-domain code search that integrates the three matching schemas.
Query-comment matching and code-code matching assume the existence of code comments and code snippet examples, respectively.
However, in practical applications, code comments are often absent and there are no feasible code snippet examples for real-time queries, hindering the use of the two matching schemas.
Inspired by the recent advancements of LLMs in zero-shot code summarization and code generation~\cite{roziere2023code,luo2023wizardcoder,li2023starcoder,achiam2023gpt}, we propose leveraging LLMs to deal with these obstacles.
Specifically, we first employ zero-shot prompting to guide LLMs in generating a comment for each code snippet in the codebase and producing a code for each query. 
Subsequently, we transform queries, code snippets, comments, and the generated code into embeddings using PLMs and assess vector similarities through three matching schemas, i.e., query-code, query-comment, and generated code-code. 
Lastly, we present a sampling-based fusion approach that combines the three similarity scores to rank the final results. 

To evaluate CodeBridge, we further investigate the following research questions:
\begin{itemize}
    \item RQ4: How does CodeBridge perform?
    \item RQ5: How effective is our fusion strategy?
    \item RQ6: How sensitive is CodeBridge to its components and hyper-parameters?
\end{itemize}
Experimental results show that in the zero-shot setting, our approach outperforms the state-of-the-art PLM-based code search approaches, i.e., CoCoSoDa~\cite{shi2023cocosoda} and UniXcoder~\cite{guo2022unixcoder}, by an average of 21.4\% and 24.9\%  in MRR, respectively, across three datasets of different domains.
Moreover, our fine-tuning-free method can yield results better than or comparable to those of RAPID~\cite{fan2024rapid}, which necessitates fine-tuning for each domain.
Further analysis demonstrates the effectiveness of our fusion strategy within our approach and reveals that our approach is effective across various retrieval models, weight selections, and LLMs.

In summary, this paper makes the following contributions:
\begin{itemize}
    \item We empirically investigate the effectiveness of query-comment matching and code-code matching in zero-shot cross-domain code search. 
    Our analysis highlights, for the first time, the high complementarity of the query-code, query-comment, and code-code matching schemas, which can be utilized to mitigate domain gaps.
    \item We propose a novel fine-tuning-free approach for zero-shot cross-domain code search.
    Our approach is easy to implement and establishes a strong baseline for zero-shot cross-domain code search.
    \item We conduct extensive experiments to evaluate the effectiveness of our approach on zero-shot cross-domain code search.
    Our approach outperforms the state-of-the-art PLM-based code search approaches and achieves performance that is better than or comparable to existing methods for zero-shot cross-domain code search that necessitate fine-tuning.
\end{itemize}

%% file: sections/related_work.tex
\section{Background and Related Work}
\subsection{Code Search}
\subsubsection{Problem Formulation}
The code search task aims to retrieve relevant code snippets from a codebase containing a set of code snippets $\{c_1,c_2,...,c_n\}$ for a natural language query $q$. 
This is accomplished by ranking the code snippets based on their computed similarity scores $\{s_1, s_2, ..., s_n\}$, where each score $s_i = f(q, c_i)$ reflects the relevance of code snippet $c_i$ to query $q$. 
The system then returns the top-$k$ code snippets as results.

In IR-based methods, $f$ is commonly a text-matching function.
In deep learning (DL)-based methods, a model first transforms $q$ and $c_i$ into vectors $v_q$ and $v_{c_i}$.
The function $f$ usually computes the cosine similarity between $v_q$ and $v_{c_i}$.
The model performing this task is called a retriever.

\subsubsection{Code Search Methods}
Code search methods can be categorized into two groups: IR-based methods and DL-based methods.
IR-based methods rely on text-matching algorithms to retrieve relevant code snippets but struggle to capture deep semantic meaning.
In contrast, DL-based methods, which utilize neural networks to learn query-code correlations in large-scale datasets, have become increasingly popular in recent years.
For example, DeepCS~\cite{gu2018deep} uses recurrent neural networks to embed queries and code in a shared vector space.

Recently, pre-trained models have demonstrated superior performance compared to traditional DL-based methods~\cite{feng2020codebert,guo2020graphcodebert,guo2022unixcoder,wang2021codet5}.
Pre-trained models first learn extensive code knowledge from large-scale code repositories and are then fine-tuned on domain-specific query-code pairs. 
To this end, many pre-trained code models have been proposed.
\cite{feng2020codebert} introduce CodeBERT, which is the first pre-trained model specifically designed for code.
\cite{guo2022unixcoder} propose UniXcoder, a unified cross-modal pre-trained model for both code-related understanding tasks and generation tasks.
\cite{shi2023cocosoda} propose CoCoSoDa, which utilizes soft data augmentation and multimodal momentum contrastive learning to align query-code pairs, achieving state-of-the-art results on the CodeSearchNet~\cite{husain2019codesearchnet} benchmark.

Our proposed approach is also built upon PLMs.
However, in contrast to the work mentioned above, this work focuses on zero-shot cross-domain code search.
In addition, our proposed approach can leverage existing models without modifying their internal structure or parameters and thus is complementary instead of competing with existing PLM-based methods.

\subsection{Cross-Domain Code Search}
Cross-domain code search refers to retrieving relevant code snippets from a target domain when the initial training data comes from different domains.
IR-based methods~\cite{zhang2021bag,lv2015codehow} utilizing unsupervised text matching algorithms can naturally handle this situation.
However, they struggle to capture deep semantics. 
DL-based methods are more effective, but adapting DL-based methods to a new domain is challenging without sufficient labeled data.
Existing methods that tackle this challenge can be divided into three categories:

{\em Pre-training} uses large-scale unlabeled data and self-supervised objectives to train neural models.
The trained model can learn common knowledge from the data and be adapted to different domains.
For example, \cite{guo2020graphcodebert} introduces edge prediction and node alignment as pre-training objectives to leverage data flow information. 
UniXcoder~\cite{guo2022unixcoder} employs a denoising objective and incorporates abstract syntax tree information in pre-training.

{\em Meta learning} uses multiple tasks to help models adapt to new domains with very
few labeled data~\cite{finn2017model}.
In code search, \cite{chai2022cross} apply Model-Agnostic Meta-Learning to improve model parameter initialization.
\cite{pian2023metatptrans} introduce MetaTPTrans, which learns language-agnostic information from multilingual source code.

{\em Pseudo-labeling} refers to generating labels for unlabeled data with existing generators and then training the model with the synthesized data.
Pseudo-labeling has been widely used in the field of natural language processing (NLP)~\cite{ma2021zero,wang2021gpl}.
\cite{fan2024rapid} are the only ones to apply pseudo-labeling to tackle the zero-shot cross-domain code search task.
They utilize pre-trained models to generate synthetic data and introduce a mixture sampling strategy to mitigate noise.
RAPID exhibits outstanding performance that surpasses all baseline models.

Unlike the methods mentioned above, our approach demands no training and requires no modification of model parameters.

\subsection{Large Language Models}
Recently, LLMs have demonstrated impressive zero-shot capabilities in diverse code-related tasks including code generation and code summarization~\cite{nijkamp2022codegen,roziere2023code,achiam2023gpt,luo2023wizardcoder,guo2024deepseek}.
{\em Code generation} refers to generating a code for a given natural language description.
In recent years, LLMs designed for coding tasks, such as CodeGen~\cite{nijkamp2022codegen}, Code Llama~\cite{roziere2023code} and DeepSeek-Coder~\cite{guo2024deepseek}, have achieved remarkable results in code generation.
{\em Code summarization} refers to generating a summary in natural language for a given code.
Many works~\cite{sun2023automatic,geng2024large} have demonstrated the effectiveness of LLMs in code summarization.

In this work, we directly utilize LLMs' zero-shot capabilities to generate code and comments for cross-domain code search.
The methods mentioned above can potentially be used in our approach.

%% file: sections/empirical_setup.tex
\section{Empirical Study Setup}
We conduct an empirical study to investigate the feasibility of leveraging query-comment matching and code-code matching to mitigate domain gaps.
In this section, we will introduce the experimental setup of our empirical study.

\subsection{Dataset}
This work focuses on cross-domain code search scenarios. 
Thus, we use the Solidity dataset~\cite{chai2022cross}, which is commonly employed as a benchmark for the cross-domain code search task~\cite{chai2022cross,fan2024rapid}.
Solidity is a language designed for smart contracts~\cite{wohrer2018smart} and is not included in the pre-trained data of the PLMs we use.
We follow the data split of prior work~\cite{chai2022cross}, where the training, validation, and test sets consist of 56,976, 4,096, and 1,000 samples, respectively.
We exclusively use the test set from the Solidity dataset, which consists of 1,000 test queries and their corresponding 1,000 Solidity functions.

\subsection{Research Questions}
Our study is structured around the following research questions (RQs):

\textbf{RQ1: How effective is query-comment matching compared to query-code matching?}
Our objective is to investigate the effectiveness of query-comment matching and the relationship between query-comment matching and query-code matching.
To answer RQ1, we first compare the performance of query-comment matching with query-code matching.
For the query-code matching schema, we calculate the similarity between the query and the code and then order the code based on the similarity scores.
For the query-comment matching schema, since the original comments of code are used as queries, we first use zero-shot prompting to guide LLMs in generating a comment for each code.
Then we order the code based on similarity scores between the query and comments related to the code. 
We then identify the differences between the retrieval results obtained from these two search methods to determine if they complement each other.
Finally, we analyze the specific scenarios in which the query-comment matching method performs better. 
    
\textbf{RQ2: How effective is code-code matching compared to query-code matching?}
Similar to RQ1, we first compare the performance of code-code matching with query-code matching.
For the query-code matching schema, the retrieval process is identical to that described in RQ1.
For the code-code matching schema, because there are no code snippets labeled as matching the test queries in either the training set or validation set, we also leverage LLMs to generate a code snippet for each query. 
During the retrieval process, we rank the code based on the similarity scores between the code and the generated code.
We then also analyze the differences between the retrieval results from the two search methods and the specific scenarios where code-code matching performs better.

\textbf{RQ3: Can the three matching schemas complement each other?}
Considering the complementarity between the query-code and query-comment schemas, as well as between the query-code and code-code schemas, we would like to investigate whether these three schemas can complement each another, and how their relationships can be leveraged for zero-shot code search.
To address this RQ, We further analyze the retrieval results from the three matching schemas and the outcome of integrating these three schemas.

\subsection{Evaluation Metrics}
We follow previous studies~\cite{fan2024rapid,cheng2022csrs,chai2022cross} and utilize two widely adopted metrics, MRR (Mean Reciprocal Rank) and top-$k$ accuracy ($k=1,5,10$), to evaluate the performance of code search approaches. 
MRR is the average of the reciprocal ranks of correct answers for a set of queries.
Top-$k$ accuracy is the proportion of queries for which relevant code can be found among the top $k$ results.

\subsection{Implementation Details}
As UniXcoder~\cite{guo2022unixcoder} has achieved remarkable results and is widely used as a backbone model in code search~\cite{shi2023cocosoda,wang2023you,wang2023one}, we utilize UniXcoder as the retriever to perform zero-shot code search.
Following previous studies~\cite{feng2020codebert,wang2023codet5+, fan2024rapid}, we set the maximum sequence length of the retriever's input to 256 for programming language (PL) and 128 for natural language (NL), respectively.
Considering both effectiveness and efficiency, we utilize DeepSeek-Coder-1.3B-Instruct for zero-shot code and comment generation due to its impressive performance and the relatively small number of parameters.
We set the maximum generation length of the LLM to 256 for PL and 128 for NL.

%% file: sections/empirical_result.tex
\section{Empirical Results}
\subsection{RQ1: How effective is query-comment matching compared to query-code matching?}
\input{tables/rq1_performance}

The results are shown in \cref{tab:paradigm}. 
The results demonstrate that query-comment matching is not superior to query-code matching. 
However, the MRR and accuracy of the query-comment approach are close to the query-code approach, suggesting promising potential for query-comment matching. 

\begin{figure}[htbp]
	\centering
	\begin{subfigure}{0.3\linewidth}
		\centering
		\includegraphics[width=0.7\linewidth]{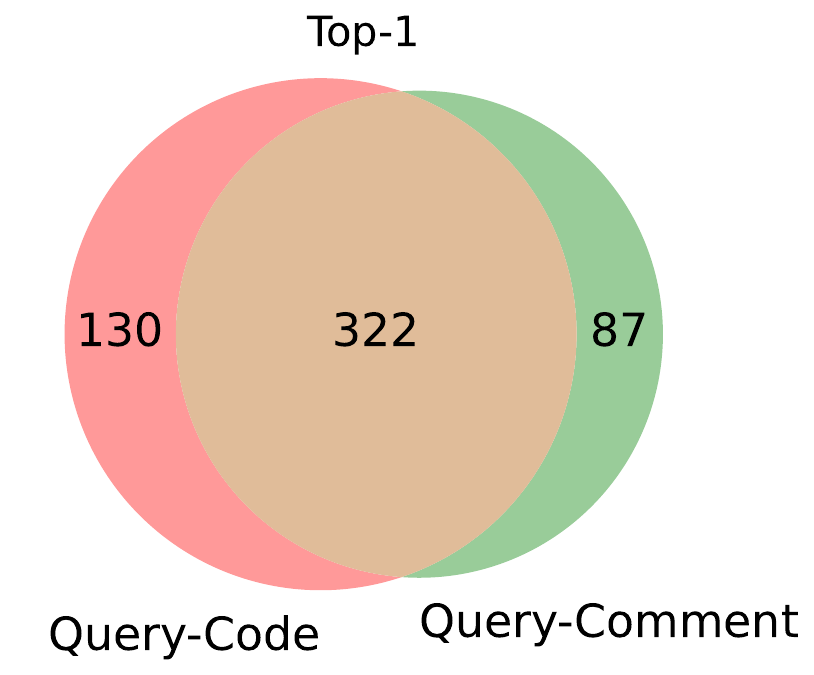}
		\caption{Query-Code vs. Query-Comment}
		\label{fig:visa}
	\end{subfigure}
	\centering
	\begin{subfigure}{0.3\linewidth}
		\centering
        \vspace{2mm}
		\includegraphics[width=0.6\linewidth]{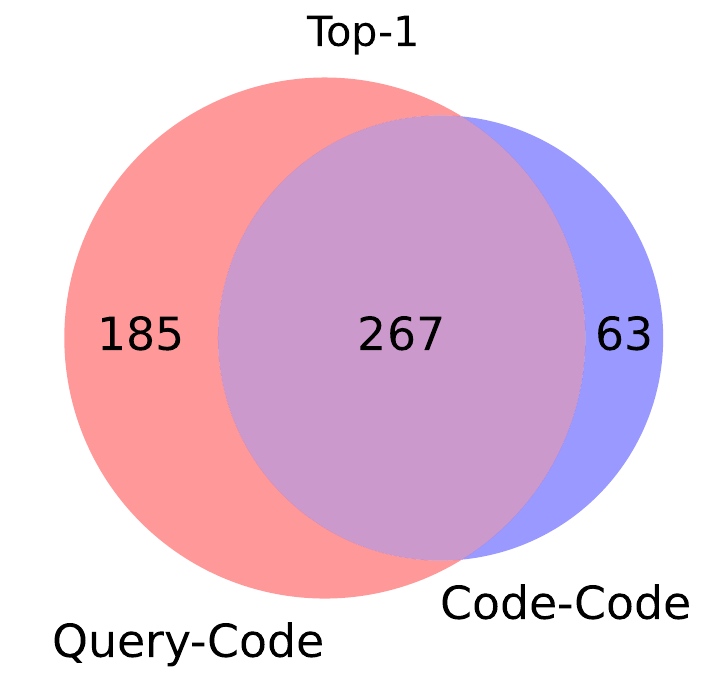}
		\caption{Query-Code vs. Code-Code}
		\label{fig:visb}
	\end{subfigure}
	\centering
	\begin{subfigure}{0.3\linewidth}
		\centering
        \vspace{2mm}
        
		\includegraphics[width=0.7\linewidth]{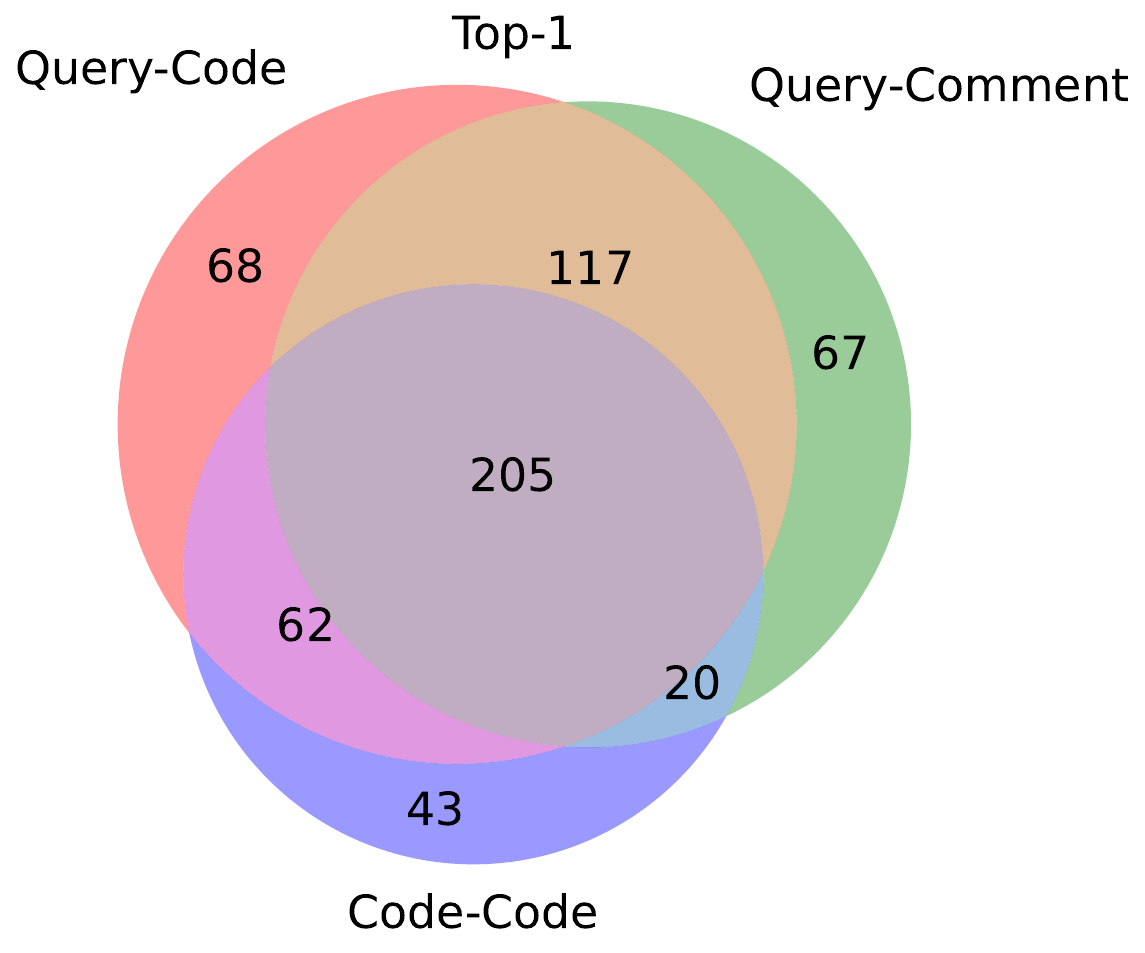}
		\caption{Three matching types}
		\label{fig:visall}
	\end{subfigure}
	\caption{The Venn diagram of the top-1 retrieved samples using UniXcoder based on the Solidity test set.}
	\label{fig:visualization}
\end{figure}

\cref{fig:visa} presents the Venn diagram of the cases where the first code snippet retrieved by the query-code approach or the query-comment approach is correct. 
The result shows that the query-comment approach successfully retrieves 87 samples that cannot be correctly handled by the query-code approach, accounting for 19.2\% of the top-1 retrieved samples of the query-comment approach.
\textbf{This indicates that the two approaches are complementary.}

\begin{figure}
    \centering
    \includegraphics[width=0.99\linewidth]{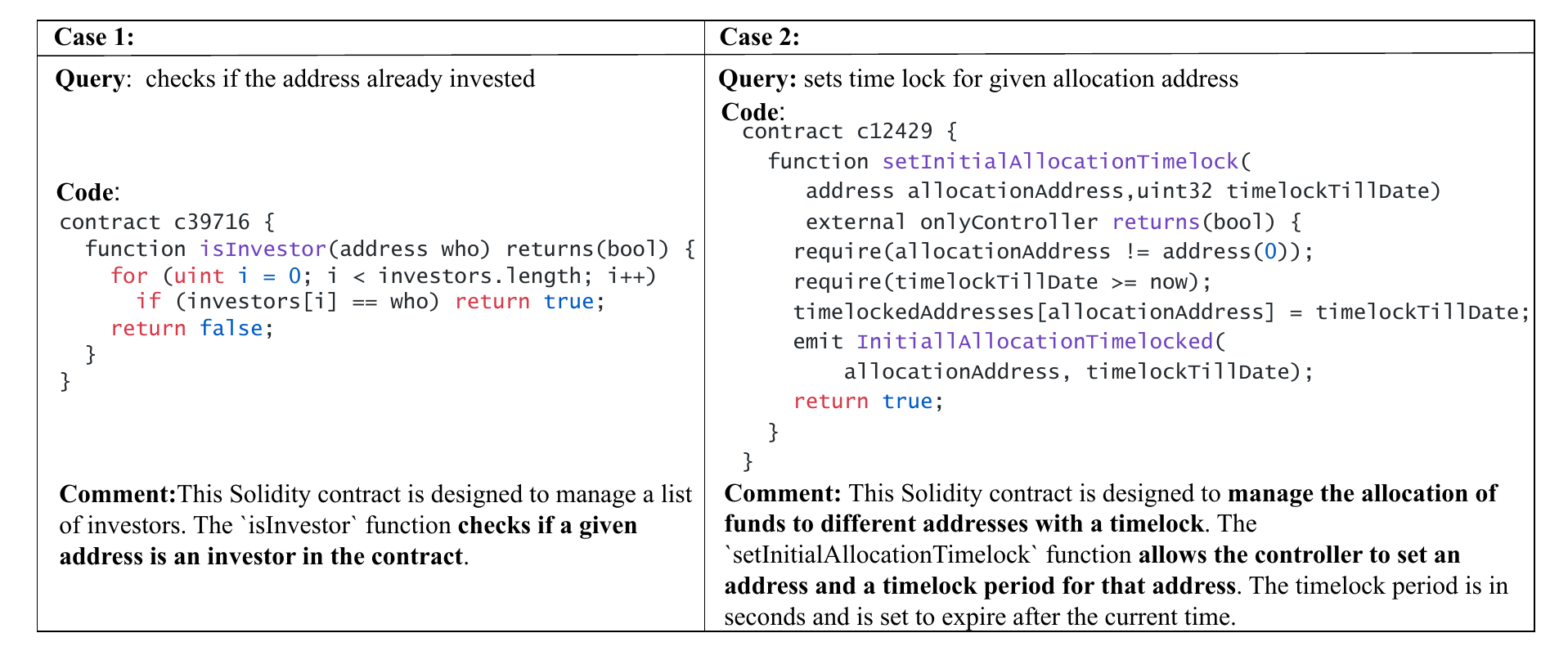}
    \caption{Cases where query-comment method outperforms query-code method in Solidity's test dataset. The code is the ground truth code for the given query. The comment is generated by LLMs.}
    \label{fig:rq1_example}
    
\end{figure}

\input{tables/rq1_proportion}

We analyze the 87 samples that are successfully handled by query-comment matching but not by query-code matching.
We observe that when the discrepancy between the query and the code is significant, and the comment effectively summarizes the function's purpose while aligning closely with the query's intent, query-comment matching outperforms query-code matching.
We further analyze the potential reasons for the better performance of query-comment matching on these samples.
We find that in contrast to query-code matching, which requires mapping knowledge between the query and the code, the comment masks the implementation details in the code and aligns with the query at the same level of abstraction.
This makes the matching process more straightforward.
\lky{We classify these potential reasons into two categories, with their respective proportions outlined in \cref{tab:proportion} and illustrative examples provided in \cref{fig:rq1_example}:}

(1) Comments mask domain-specific implementation details.
In Case 1 illustrated in \cref{fig:rq1_example}, the query-code matching model faces challenges in correlating the query  ``checks if the address already invested'' with specific details such as the variable type \textit{address}, the concept of investors, and the equality comparison \textit{investors[i] == who} of type \textit{address}.
However, the comment abstracts these details and clearly states the function's intent, thereby simplifying the comparison process.

(2) Comments mask unnecessary details. 
In Case 2 depicted in \cref{fig:rq1_example}, the target code matches the query ``sets time lock for given allocation address''. 
However, the code also involves additional steps like validating the address, which are not directly relevant to the primary action described by the query. Such extra details can introduce noise when aligning queries with code. 
In contrast, comments concisely summarize the core functionality, reducing noise and enabling accurate matching.

By analyzing the 130 samples handled successfully by query-code matching but not query-
comment matching, we observe two failure causes for query-comment matching, i.e., (1) imprecise comments and (2) excessive noise. Imprecise comments hinder the alignment between queries and comments. For example, a comment describes the code as a ``CryptoKitties-like game token contract'' while omitting its key function of initiating contributions. This naturally leads to a mismatch with the query ``Start Contribute''. Meanwhile, excessive noise drowns out key information, weakening the match. For example, a comment accurately describes the target function ``is used to add a new owner to the contract'' but also includes extraneous details, such as the usage of modifiers. This additional information weakens the match between ``add a new owner'' and the query ``Adds an owner''. These findings highlight that only relying on query-comment matching is not enough for accurate code search.

\rqbox{
\textbf{Answer to RQ1}: When used individually, query-comment matching does not outperform query-code matching. 
However, they are complementary. 
Further analysis suggests that query-comment matching can perform better than query-code matching in certain scenarios.
This may be because comments often omit domain-specific and/or unnecessary details in the code, reducing the need for domain-specific query-code mapping knowledge.
\lky{Query-comment matching might fail when comments are imprecise or contain excessive noise.}
}\label{finding11}

\subsection{RQ2: How effective is code-code matching compared to query-code matching?}

The experimental results are shown in \cref{tab:paradigm}.
The results demonstrate that code-code matching is also not superior to query-code matching.
\cref{fig:visb} presents the Venn diagram of the cases where the first code snippet retrieved by the query-code approach or the code-code approach is correct.
The result shows that the code-code matching successfully retrieves 63 samples that cannot be correctly handled by the query-code approach, accounting for 13.9\% of the top-1 retrieved samples of the query-code approach.
\textbf{This also indicates that the two approaches are complementary.}

\begin{figure}[!htbp]
    \centering
    \includegraphics[width=1\linewidth]{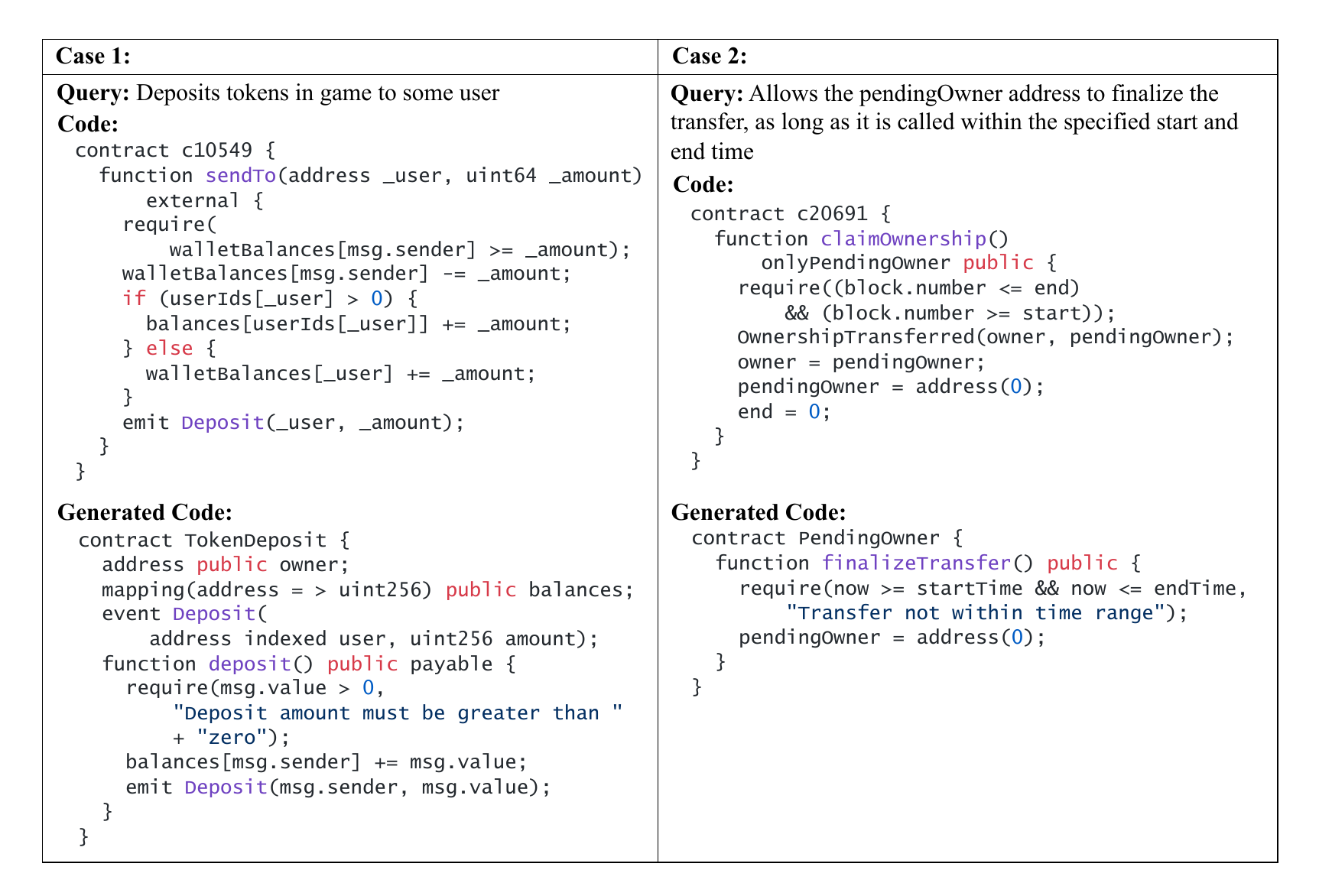}
    \caption{Cases where code-code method outperforms query-code method in Solidity's test dataset. The code represents the correct code for the query, while the generated code denotes the code generated by LLM for the same query.}
    \label{fig:rq2_example}
\end{figure}

Similar to RQ1, we analyze the 63 samples that are successfully handled by the code-code matching but not by the query-code matching.
We find that in these cases, the generated code shares a similar implementation to the target code.
The reason why code-code matching outperforms query-code matching in these samples may be attributed to the fact that the two code snippets operate within the same abstract level.
Specifically, code deals with low-level operations, and directly comparing these low-level implementations can help reduce the need for domain-specific query-code mapping knowledge.
\lky{We classify these potential reasons into two categories, with their respective proportions outlined in \cref{tab:proportion} and illustrative examples provided in \cref{fig:rq2_example}:}

(1) In case 1 illustrated in \cref{fig:rq2_example}, both the generated code and the target code start a deposit by using ``emit Deposit'', a specific syntax in Solidity. 
Such direct match reduces the difficulty for the model to identify the relevant answer. 
In contrast, the query-code matching schema fails in this case because the operation of depositing tokens is relatively rare in other programming languages like Python and understanding how to deposit tokens can be difficult without domain-specific knowledge.
This case indicates that the implementation similarities between the generated code and the target code can facilitate a more direct match and reduce the difficulties in cross-domain code search.

(2) In case 2 illustrated in Figure \ref{fig:rq2_example}, both the generated code and the target code implement time constraints by using \textit{require} statement and a mechanism to finalize the ownership transfer in the same order. 
The model can identify the comparable operation sequences directly without first learning the code's structure for a domain-specific query.
However, the query-code matching schema fails in this case because the specific implementation order of operations for finalizing a transfer is difficult to determine without domain-specific knowledge.
This case indicates that structure similarities between the generated code and the target code can also help reduce the need for domain-specific knowledge.

\lky{
We analyze the 185 samples successfully handled by query-code matching but not by code-code matching. We identify two failure causes for code-code matching: (1) imprecise generated code and (2) differing implementations. Imprecise code often leads to mismatches. For example, the generated code that simply returns a \textit{byte32} variable naturally fails to match the query ``Extract 256-bit worth of data from the bytes stream''. Differing implementations also hinder code-code matching. For example, for the query ``create a new offer with setting'', the generated code uses a struct \textit{Offer} with an initialization method, while the target code uses an internal function \textit{CreateOffer\_internal}. This implementation mismatch prevents successful code-code matching.}

Results in \cref{tab:paradigm} show that code-code matching underperforms compared to query-comment matching.
This may be due to the diversity of code: developers use varying identifiers, and implementations differ in abstraction levels, such as low-level instructions versus high-level library functions. 
These factors make code-code matching harder than simpler query-comment matching.

\rqbox{
\textbf{Answer to RQ2:}
When used individually, code-code matching does not surpass query-code matching in performance.
However, the two schemas are also complementary.
Further analysis indicates that code-code matching might outperform query-code matching in certain scenarios. 
The main reason is that code-code matching focuses on the same low-level operations, which reduces the need for domain-specific knowledge to map queries to code.
\lky{Code-code matching might fail when the generated code is imprecise or using a different implementation.}
}\label{finding2}
\subsection{RQ3: Can the three matching schemas complement each other?}
\lky{
The three matching schemas each have distinct advantages.
Query-comment matching simplifies the matching process by masking implementation details.
Code-code matching compares low-level implementations in the code, reducing the need for domain-specific knowledge.
We further analyze the advantages of query-code matching by examining: (1) 130 samples successfully handled by query-code matching but not by query-comment matching; (2) 185 samples successfully handled by query-code matching but not by code-code matching.
We also identify two specific scenarios where query-code matching outperforms the other matching schemas:}

\lky{(1) \textbf{Direct function name matching}: Docstring-code pairs, where function names usually align with docstrings, are used to train code search models like CoCoSoDa.
Thus, query-code matching effectively identifies relevant code snippets when the query directly matches a function name.
}

\lky{(2) \textbf{Function specification matching}: Query-code matching performs better when the query includes function specifications (e.g., return types, input parameters), which can be directly mapped to the code implementation for accurate matching.}

\cref{fig:visall} presents the Venn diagram of samples from the three matching schemas where the correct code is successfully ranked first.
The three matching schemas exhibit high complementarity.
The query-comment matching schema retrieves 67 samples not found in the query-code and code-code retrieval samples.
Meanwhile, the code-code matching schema retrieves 43 samples that are not included in the set of query-code and query-comment retrieval samples. 
By combining the top-1 retrieval samples from the three matching schemas, an additional 130 samples can be retrieved at the top rank, constituting 28.8\% of the top-1 retrieval samples from the query-code matching schema.
This underscores the significant potential of integrating these three matching schemas to enhance code search performance.

\rqbox{
\textbf{Answer to RQ3:}
The three matching schemas are complementary.
Combining retrieval results of three matching approaches can yield a significant 28.8\% enhancement in top-1 accuracy over query-to-code matching alone.
}\label{finding3}

%% file: tables/rq1_performance.tex
\begin{table}[!htbp]
  \centering
  \belowrulesep=0pt
  \aboverulesep=0pt
  \renewcommand\arraystretch{1.2}
  \setlength\tabcolsep{2pt} 
  \scriptsize
  \caption{Performance on Solidity with Different Matching Schemas}
  \label{tab:paradigm}
  \begin{tabularx}{10cm}{C{2.5cm}|C{1.7cm}|C{1.7cm}|C{1.7cm}|C{1.7cm}}
    \toprule
    Schema & MRR & Top-1 & Top-5 & Top-10 \\
    \midrule
    Query-Code &  0.544   & 0.452 & 0.651 &  0.701 \\
Query-Comment  & 0.500 & 0.409 & 0.605 & 0.681 \\
Code-Code   &  0.410   &  0.330 & 0.492 & 0.566 \\
  \bottomrule
\end{tabularx}
\end{table}

%% file: tables/rq1_proportion.tex
\begin{table}[H]
  \centering
  \belowrulesep=0pt
  \aboverulesep=0pt
  \renewcommand\arraystretch{1.2}
  \color{\tablecolor}
  \scriptsize
  \setlength\tabcolsep{2pt} 
  \caption{\small \lky{Proportions of categories}}
  
  \label{tab:proportion}
  \begin{tabularx}{12.5cm}{C{2.8cm}|C{1.2cm}|C{6.6cm}|C{1.3cm}}
  \toprule
  Outperforming Pattern & \#Samples  & Category & Proportion \\
  \midrule
  \multirow{2}{*}{Query-Comment } & \multirow{2}{*}{87}
  & Comments mask domain-specific implementation details & 62.0\% \\
  \cline{3-4}
  && Comments mask unnecessary details & 38.0\% \\
  \midrule
  \multirow{2}{*}{Code-Code} & \multirow{2}{*}{63}
  & Direct token matching  & 28.6\% \\
  \cline{3-4}
  
  && Operation sequence matching & 71.4\% \\
  \bottomrule
\end{tabularx}
\end{table}

%% file: sections/approach.tex
\section{Approach of CodeBridge}
The results of our empirical study show that query-comment matching can hide domain-specific and unnecessary details in the code, thereby enabling matching at a similar level of abstraction. 
Meanwhile, code-code matching facilitates a direct comparison of the concrete implementation details, thus focusing the matching on lower-level operations. 
Each method provides distinct advantages in specific cross-domain scenarios and can serve as a complement to query-code matching.
Inspired by these findings, we propose a novel method named CodeBridge to bridge domain gaps for cross-domain code search by integrating the three matching schemas.

\begin{figure}[!htbp]

    \centering
    \includegraphics[width=0.9\linewidth]{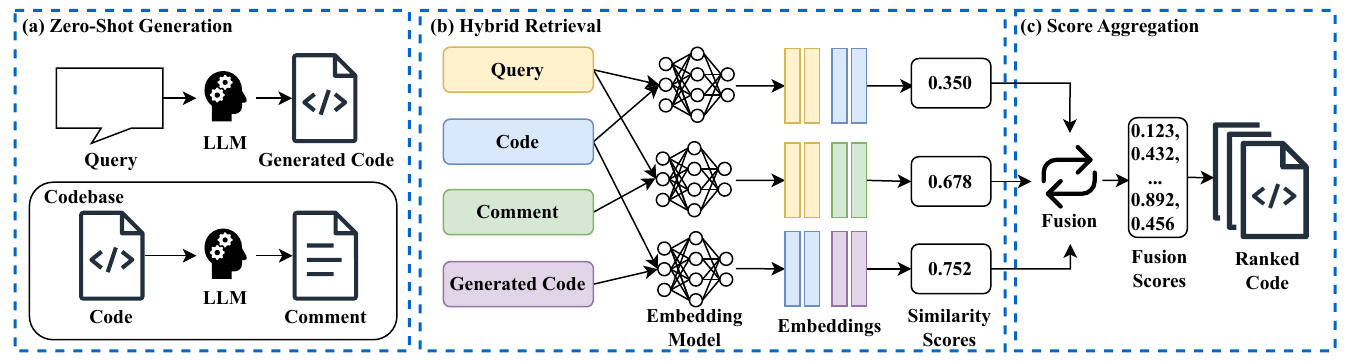}
    \caption{Overall Framework of CodeBridge}
    \label{fig:framework}

\end{figure}

\cref{fig:framework} shows the overall framework of CodeBridge.
CodeBridge is composed of three components: \textbf{zero-shot generation}, \textbf{hybrid retrieval}, and \textbf{score aggregation}.
(1) In the zero-shot generation stage, queries are translated into code snippets, and conversely, code is translated into descriptive comments. 
As comments are often missing in practical applications, and the code snippets that satisfy real-time queries are also unknown during runtime, we use zero-shot prompting to guide LLMs in generating code and creating code summaries.
(2) In the hybrid retrieval stage, all entities, including queries, code, comments, and generated code, are transformed into vector representations through embedding models. 
We then calculate the similarity scores for each matching pair.
(3) In the score aggregation stage, we combine the scores and rank them to determine the final result.
We propose a sampling-based fusion strategy to aggregate the scores.
\subsection{Zero-Shot Generation}
In this stage, we use zero-shot prompting to guide LLMs in generating comments and code.
We design prompts, as presented in \cref{tab:prompt}, for code summarization and code generation.
For code summarization, we take inspiration from \cite{sun2023automatic} to create our prompts. 
As shown in \cref{tab:prompt}, we include details about the programming language to help the LLM understand and tap into relevant expertise. 
Also, we take out the phrase ``in one sentence'' that is used in Sun et al.'s study~\cite{sun2023automatic} to make the summaries richer and more informative.
For code generation, we request the LLMs to generate code in a specific language and ensure that it provides an answer.

\begin{table}[!htbp]
    \centering
    \footnotesize
    \renewcommand\arraystretch{1.2}
  \setlength\tabcolsep{2pt} 
    \belowrulesep=0pt
  \aboverulesep=0pt
    \caption{Prompts for Code Summarization and Code Generation}
  
    \begin{tabular}{m{2.4cm}<{\centering}|m{10cm}<{\raggedright}}
    \toprule
    Task & \makecell[c]{Prompt}\\
    \midrule
         Code Summarization & Below is a \{language\} code that describes a task. Please give a short summary describing the purpose of the code. You must write only summary without any prefix or suffix explanations. \textbackslash n
    \{code\} \\
    \midrule
      Code Generation & Write a code for the following query in \{language\} without comments. You must return a code and must not refuse to answer. \textbackslash n
    \{query\} \\
    \bottomrule
    \end{tabular}
    \label{tab:prompt}
\end{table}

\subsection{Hybrid Retrieval}
In this stage, we calculate similarity scores in three matching schemas.
After zero-shot generation, the initial query-code pair, denoted as $\langle{}q,c\rangle{}$ where $q$ represents the query and $c$ the source code, is expanded to $\langle{}q,c,m,g\rangle{}$. 
Here, $m$ represents the comment associated with the code $c$, and $g$ represents the code generated from the query $q$.
The three matching schemas are defined as follows:
\begin{itemize}
    \item \textbf{Query-Code Matching ($\langle{}q,c\rangle{}$)}: Direct comparison of the query against the source code. 
    \item \textbf{Query-Comment Matching ($\langle{}q,m\rangle{}$)}: 
    Comparison at an abstract level between the query and the descriptive comment.
    \item \textbf{Code-Code Matching ($\langle{}c,g\rangle{}$)}: Evaluating the implementation consistency between the generated code snippet ($g$) and the original code ($c$).
\end{itemize}
Let $\phi(x, y)$ denote the similarity scoring function, where $x$ and $y$ represent the elements being compared.
We first encode input into vectors using a PLM, denoted as $\mathcal{M}$. Then we calculate the similarity between vectors.
We use cosine similarity by default. The formulas are as follows.
\begin{equation}
v_x = \mathcal{M}(x),\quad
v_y = \mathcal{M}(y),\quad
\phi(x,y) = sim(\mathcal{M}(x),\mathcal{M}(y))
\end{equation}

For each expanded $\langle{}q,c,m,g\rangle{}$ tuple, the similarity scores are computed as follows:
\begin{equation}
    s_{qc} = \phi_{qc}(q,c),\quad
    s_{qm} = \phi_{qm}(q,m),\quad
    s_{cg} = \phi_{cg}(c,g)
\end{equation}
where $s_{\text{type}}$ represents the similarity score for the corresponding type of matching schema.

\subsection{Score Aggregation}
\label{sec:sample}
In this stage, we aim to aggregate the three similarity scores and rank the code snippets to obtain the final result.
The most straightforward method for combining different retrieval outputs is a linear combination.
However, the assignment of weights is influenced by two key aspects: the comparative matching ability of different models and the varying quality of both generated comments and synthesized code. 
This interplay introduces complexity in determining the ideal weights.

To obtain an appropriate weight configuration for a specific ensemble of models, we propose a sampling-based linear combination method. 
Consider the equation for computing the final score:
\begin{equation}
    \label{eq:linear}
   s_{total} = \alpha \times s_{qc} + \beta \times s_{qm} + \gamma \times s_{cg}
\end{equation}
To determine suitable values for $\alpha$, $\beta$, and $\gamma$, we first randomly select 1,000 instances from the training set of CodeSearchNet-Java~\cite{husain2019codesearchnet}, a widely-used dataset for code search, to serve as the validation set.
These instances are all in Java, a language not represented in our evaluation datasets.
We then use grid search with a 0.05 step size to explore the full range of potential weight combinations.
The configuration yielding the maximum top-10 accuracy is adopted.

%% file: sections/rq.tex
\section{Evaluation Settings}
We conduct comprehensive experiments to evaluate the performance of CodeBridge.
In this section, we first introduce the research questions.
Then we describe the datasets, baselines, evaluation metrics, and implementation details.
\subsection{Research Questions}
We systematically evaluate the effectiveness (RQ4 and RQ5) of CodeBridge and analyze its sensitivity to its components and hyper-parameters (RQ6).

\textbf{RQ4: How does CodeBridge perform?}
We evaluate CodeBridge across three datasets of different domains against various code search methods, including the state-of-the-art PLM-based code search approaches such as UniXcoder~\cite{guo2022unixcoder} and CoCoSoDa~\cite{shi2023cocosoda}, as well as the state-of-the-art zero-shot cross-domain code search method, RAPID~\cite{fan2024rapid}.
As LLMs might directly serve as in-domain code search tools, we implement zero-shot LLM embedding methods for code search and evaluate their performance against CodeBridge.
Additionally, we perform an ablation study to evaluate the effectiveness of combining three matching schemas.

\textbf{RQ5: How effective is our fusion strategy?}
Existing fusion strategies can be divided into two groups: score-based (e.g., CombSUM~\cite{fox1994combination} and CombMNZ~\cite{fox1994combination}) and rank-based (e.g., Borda~\cite{borda1784memoire}, RRF~\cite{cormack2009reciprocal}).
We compare our strategy with four widely-used fusion strategies~\cite{benham2017risk} on the Solidity test set.

\textbf{RQ6: How sensitive is CodeBridge to its components and hyper-parameters?}
To analyze the sensitivity of CodeBridge, we conduct experiments on CodeBridge with different retrieval models, weight selections, and LLMs:

(1) \textit{Different retrieval models}:
we evaluate our framework with different retrieval models and their combinations on three datasets. 

(2) \textit{Different weight selections}:
We investigate the effects of varying weights in \cref{eq:linear} and the impact of the sampling dataset.
First, we utilize the Solidity dataset to systematically explore the performance of all possible weight configurations under the constraint that $\alpha + \beta +\gamma = 1$, using a step size of 0.01.
We present the findings through visual representations.
Second, we obtain weights from datasets of three other widely used languages in CodeSearchNet~\cite{husain2019codesearchnet}.
Then, we analyze the impact of the sampling dataset.

(3) \textit{Different LLMs}:
We use various LLMs to perform zero-shot generation and test our approach on the Solidity dataset.

\subsection{Datasets} 
Following previous work~\cite{fan2024rapid}, we conduct experiments on three datasets: SQL~\cite{chai2022cross}, Solidity~\cite{chai2022cross} and CoSQA~\cite{huang2021cosqa}.
Each dataset includes a set of test queries and a codebase.
The SQL dataset and Solidity dataset contain code examples written in SQL and Solidity, respectively.
These languages are not part of the training dataset for the PLMs we use.
Instead of using code comments as queries, the CoSQA dataset includes real queries from Microsoft Bing~\cite{bingPreservingParadise} and corresponding Python code snippets from Github~\cite{githubGitHubLets}.

The statistics of the three datasets are shown in \cref{tab:datasets}.
Recognizing that code comments are often missing or insufficient in practice, we remove all code comments in the code. 
Besides, we utilize real web queries as test queries within the CoSQA dataset following the initial setting of \cite{huang2021cosqa}. 
Please note that the dataset configuration of RAPID, the state-of-the-art zero-shot cross-domain code search approach, is different. 
They do not fully remove code comments, and they use code comments as test queries in the CoSQA dataset.
Additionally, RAPID applies a custom split to the CoSQA dataset, resulting in different dataset sizes. The CoSQA dataset consists of 20,604 queries and 6,267 code snippets, where each code snippet is accompanied by a comment. RAPID constructs three subsets: 20,000 code-only samples for training (including repeated code snippets), 602 comment-code pairs for validation, and 901 comment-code pairs for testing. RAPID generates comments for each code snippet in the training set to create the final comment-code pairs for training.
To ensure a fair comparison, when comparing our approach to RAPID, we use the dataset configuration of RAPID.
Importantly, the difference in dataset sizes does not affect the validity of our results, as we only compare CodeBridge and the baselines under the same setting, never cross settings.

\begin{table}[!htbp]
\vspace{-5mm}

  \centering
  \belowrulesep=0pt
  \aboverulesep=0pt
  \renewcommand\arraystretch{1.2}
  \footnotesize
  
  \caption{Statistics of the Datasets}
  \setlength\tabcolsep{2pt} 
  \label{tab:datasets}
  \begin{tabularx}{12cm}{>{\centering\arraybackslash}X|*{7}{c}}
  \toprule
  Setting & Language & Train & Valid & Test & Codebase & w/ Comment & Use Web Query \\
  \midrule
  \multirow{3}{*}{Cross-Domain Setting} 
  & SQL & 14,000 & 2,068 & 1,000 & 1,000  & no &/  \\
  & Solidity & 56,976& 4,096& 1,000 & 1,000 & no & / \\
  & CoSQA & 19,604 & 500 & 500 & 6,267 & no & yes \\
  \midrule
  \multirow{3}{*}{RAPID's Setting} 
  & SQL & 14,000 & 2,068 & 1,000 & 1,000 & no &/  \\
  & Solidity &  56,976& 4,096& 1,000 & 1,000 & yes & / \\
  & CoSQA & 20,000 & 602 & 901 & 901 & no & no \\
  
\bottomrule
\end{tabularx}
\vspace{-5mm}

\end{table}

\subsection{Baselines}
To comprehensively evaluate the performance of our approach, we broadly select traditional IR-based models, pre-trained models, zero-shot cross-domain code search methods, and zero-shot LLM embedding methods as baselines:

\textbf{IR-based Models}:
    \textbf{BM25~\cite{robertson2009probabilistic}} is an enhanced text-matching algorithm based on TF-IDF, serving as a strong baseline for unsupervised code search. 

\textbf{Pre-trained Models}:
    \textbf{GraphCodeBERT~\cite{guo2020graphcodebert}} is a code pre-trained model that leverages control flow graph information during pre-training;
        \textbf{CodeT5+ 110M~\cite{wang2023codet5+}} incorporates tasks such as causal language modeling and text-code contrastive learning to improve embedding quality;
        \textbf{UnixCoder~\cite{guo2022unixcoder}} is trained using ASTs (Abstract Syntax Trees) and code comment data, enhancing its code understanding ability.
        \textbf{CoCoSoDa~\cite{shi2023cocosoda}} applies multimodal momentum contrastive learning and soft data augmentation to enhance code and query representations.
        It achieves state-of-the-art performance on the CodeSearchNet dataset.
    
\textbf{Zero-Shot Cross-Domain Methods}:
   \textbf{RAPID~\cite{fan2024rapid}} uses a pre-trained generative model to generate queries for each code snippet and then fine-tunes the retrieval model using these synthesized query-code pairs. 
        This approach achieves state-of-the-art performance in the scenario of zero-shot cross-domain code search.

\textbf{Zero-Shot LLM Embedding Methods}:
        \textbf{Weighted Mean Pooling~\cite{muennighoff2022sgpt}} computes the positional weighted average of token embeddings;
        \textbf{Mean Pooling~\cite{reimers2019sentence}} computes the average of token embeddings;
        \textbf{EOS Pooling\cite{reimers2019sentence}} uses the last token's embedding;
        \textbf{Echo Embedding~\cite{springer2024repetition}} is the state-of-the-art zero-shot embedding method for LLMs, which repeats the input twice for richer contextual embeddings.

\subsection{Evaluation Metrics and Implementation Details}

We follow previous RQs and utilize MRR and top-$k$ accuracy ($k=1,5,10$) as the evaluation metrics.

We employ the most advanced models tailored for each type of matching schema in our approach.
Specifically, we employ CoCoSoDa for query-code, BGE~\cite{xiao2024c} for query-comment, and UniXcoder for code-code matching.
BGE is a superior natural language vector model trained on 300 million pairs of natural language texts, demonstrating powerful natural language representation capabilities. 
Specifically, we utilize the bge-large-en-v1.5 (326M).
The weights in \cref{eq:linear} of this particular model configuration are $\alpha=0.65,\beta=0.25,\gamma=0.10$.
We set the maximum sequence length of the retriever's input to 256 for PL and 128 for NL, respectively.

We use DeepSeek-Coder-1.3B-Instruct for zero-shot generation and set the maximum generation length to 256 for PL and 128 for NL.
For RAPID, we directly copy its results from its original paper~\cite{fan2024rapid}.
For PLM-based methods, we use their official implementation.
As IR-based methods are sensitive to pre-possessing methods, we follow \cite{zhang2021bag} for BM25 preprocessing steps.
We apply LLM embedding methods directly to DeepSeek-Coder-1.3B-Instruct, generating query and code embeddings for code search.

%% file: sections/rq_result.tex
\section{Evaluation Results}
\subsection{RQ4: How does CodeBridge perform?}
\label{sec:performance}
The experimental results are shown in \cref{tab:our-setting} and \cref{tab:rapid}, which demonstrate CodeBridge's effectiveness in all the evaluated languages and settings.

\input{tables/rq4_our}

According to Table~\ref{tab:our-setting}, CodeBridge surpasses CoCoSoDa by an average of 21.4\% in MRR and 30.6\% in top-1 accuracy. 
Additionally, our approach outperforms UniXcoder with an average improvement of 24.9\% in MRR and 35.4\% in top-1 accuracy.
While CoCoSoDa performs well across most languages, it doesn't work as effectively on SQL dataset, which indicates its limitation on generalization to different domains.
\lky{Meanwhile, all zero-shot LLM embedding methods perform poorly.
This may be because the unidirectional causal mask in LLMs hinders bidirectional understanding, and unlike natural language texts, source code is highly structured, which may make the text embedding method not well-suited for code search.
}
In contrast, CodeBridge exhibits consistent and superior performance across the three datasets, highlighting its great cross-domain code search capabilities.

\input{tables/rq4_rapid}

We compare CodeBridge with RAPID under its setting, as shown in Table~\ref{tab:rapid}.
The performance of CodeBridge is comparable to that of RAPID on the Solidity and CoSQA datasets. 
CodeBridge even outperforms CoCoSoDa with RAPID by 3.3\% in MRR and 3.3\% in top-1 accuracy on the SQL dataset. 
It is noteworthy that while RAPID requires collecting code snippets in the target domain and fine-tuning with synthesized data, while \textbf{our approach demands no training, and requires no intervention on the model, and thus is easy to implement and use.}
Thus, we believe our approach provides unique benefits for zero-shot cross-domain code search.

\lky{
We also conduct an ablation study to compare the performance of the three matching schemas. As shown in Table~\ref{tab:our-setting}, the results indicate that CodeBridge, which integrates all three schemas, outperforms each matching schema by substantial margins. Specifically, the average improvements of CodeBridge over query-code, query-comment, and code-code matching in MRR are 21.4\%, 15.9\%, and 52.1\%, respectively, across the three datasets. These results suggest the effectiveness of fusing the three matching schemas, consistent with our empirical findings.}

\rqbox{
\textbf{Answer to RQ4}: CodeBridge significantly outperforms existing zero-shot code search baselines by 21.4\% to 24.9\% on average in terms of MRR.
Additionally, our fine-tuning-free approach yields results that are comparable to or even better than the state-of-the-art zero-shot cross-domain code search method RAPID, which requires fine-tuning.
\lky{
The ablation study confirms the effectiveness of fusing the three matching schemas.
}
}\label{finding4}

\subsection{RQ5: How effective is our fusion strategy?}

\begin{table}[!htbp]
  \centering
  \belowrulesep=0pt
  \aboverulesep=0pt
  \renewcommand\arraystretch{1.2}
    \scriptsize
  \caption{Results of different strategies on Solidity dataset. Since CombSUM and CombMNZ calculate the frequency of occurrences in different result lists, they will reduce to an equally weighted linear combination when the recall size equals the total size of the dataset. Thus, set a recall of 10 for CombSUM and CombMNZ.}
  \setlength\tabcolsep{2pt} 
  \label{tab:fusion-strategy}
  \begin{tabularx}{10.2cm}{C{1.8cm}|C{1.5cm}|C{1.5cm}|C{1.5cm}|C{1.5cm}|C{1.5cm}}
  \toprule
  Fusion Strategy & CombSUM & CombMNZ & RRF & Borda & Our Strategy \\
  \midrule
  Top-1 & 0.523 & 0.523 & 0.494 & 0.480 & \textbf{0.571}\\
  Top-5 & 0.748 & 0.748 & 0.705 & 0.655 & \textbf{0.762} \\
  Top-10 &0.803 & 0.803 & 0.799 & 0.727 & \textbf{0.833} \\

\bottomrule
\end{tabularx}
\end{table}

\cref{tab:fusion-strategy} presents the performance of our fusion strategy.
While score-based methods outperform rank-based methods, our strategy outperforms the four widely-used fusion strategies by substantial margins, demonstrating the effectiveness of our fusion strategy.
Specifically, our strategy outperforms the best-performing baselines, i.e., CombSUM and CombMNZ, by 9.2\% in terms of top-1 accuracy. 
These results can be attributed to our strategy's comprehensive consideration of both the model's semantic understanding ability and the quality of generated content.
\rqbox{
\textbf{Answer to RQ5}: The proposed fusion strategy outperforms four widely-used fusion strategies by substantial margins.
}\label{finding5}

\subsection{RQ6: How sensitive is CodeBridge to its components and hyper-parameters?}
\subsubsection{Different Retrieval Models}

\begin{table}[!htbp]
  \centering
  
    \belowrulesep=0pt
    \aboverulesep=0pt
  \caption{Performance comparison with different retrieval models.
    "Uni" denotes UniXcoder, and "Co" denotes CoCoSoDa. 
    Each component in the combination of models is sequentially used for query-code matching, query-comment matching, and code-code matching. 
    CodeBridge integrates UniXcoder, BGE, and CoCoSoDa. The best result is bolded, and the second best is underlined.}
  \scriptsize
  \setlength\tabcolsep{3pt}
  \renewcommand\arraystretch{1.2}
  \label{tab:retrieval-model}
  \begin{tabularx}{\textwidth}{
  >{\centering\arraybackslash\hsize=1.2\hsize}X
  |*{4}{>{\centering\arraybackslash\hsize=0.5\hsize}X}
  |*{4}{>{\centering\arraybackslash\hsize=0.5\hsize}X}
  |*{4}{>{\centering\arraybackslash\hsize=0.5\hsize}X}}

\toprule
\multirow{2}{*}{\textbf{Model}}
& \multicolumn{4}{c|}{\textbf{SQL}} 
& \multicolumn{4}{c|}{\textbf{Solidity}} 
& \multicolumn{4}{c}{\textbf{CoSQA}}   \\

&MRR&Top-1&Top-5&Top-10&MRR&Top-1&Top-5&Top-10&MRR&Top-1&Top-5&Top-10 \\
\midrule
UniXcoder& 0.744 & 0.632 &0.887& 0.938 &0.544 &0.452 &0.651& 0.701 &0.376& 0.256& 0.512& 0.620 \\
Uni-Uni-Uni& 0.795& 0.697& 0.914& 0.957& 0.581& 0.491& 0.687 &0.737 &0.426& 0.296&0.572& 0.696 \\
Uni-BGE-Uni& \textbf{0.824} &\textbf{0.737}& \textbf{0.940} &\textbf{0.973}&0.640 &0.559& 0.736& 0.800&0.513& 0.382&0.654& 0.776 \\
\midrule
CoCoSoDa& 0.555& 0.441&0.690&0.769&0.625&0.541&0.728&0.788&0.482&0.346&0.630&0.724 \\
Co-Co-Co& 0.592&0.478&0.732&0.804&0.635&0.547&0.747&0.794&0.497&0.364&0.644&0.744 \\
Co-BGE-Co&0.779& 0.687& 0.899& 0.934& \underline{0.657}& \textbf{0.572}& \textbf{0.765}& \underline{0.820}&\underline{0.525}& \underline{0.390}& \underline{0.676}& \underline{0.772} \\
\midrule
CodeBridge & \underline{0.811}&\underline{0.723}&\underline{0.929}&\underline{0.958}&\textbf{0.658}&\underline{0.571}&\underline{0.762}&\textbf{0.833}&\textbf{0.544}&\textbf{0.424}&\textbf{0.682}&\textbf{0.782} \\
\bottomrule
\end{tabularx}
\end{table}

Based on the results shown in \cref{tab:retrieval-model}, we find that:
(1) \textbf{Integrating tailored models for each matching schema contributes significantly to the effectiveness of our approach}.
Specifically, models incorporating BGE (i.e., Uni-BGE-Uni and Co-BGE-Co) surpass those without BGE (i.e., Uni-Uni-Uni and Co-Co-Co).
For example, Uni-BGE-Uni outperforms Uni-Uni-Uni by an average of 11.4\% in MRR and Co-BGE-Co outperforms Co-Co-Co by an average of 13.6\% in MRR.
This can be attributed to the superior natural language understanding ability of NL embedding models, which provides more accurate query-comment matching than code pre-trained models.
CodeBridge also surpasses Uni-BGE-Uni on two datasets, highlighting the advantage of using an advanced code search model like CoCoSoDa for query-code matching.
In addition, CodeBridge's performance is better than or comparable to Co-BGE-Co, indicating that using UniXcoder, which performs outstandingly in code-to-code search tasks, contributes to the effectiveness of our method.
(2) \textbf{Our approach exhibits consistent performance across different retrieval models}. 
All variants outperform their corresponding base models, i.e., UniXCoder and CoCoSoDa, by substantial 
s.
For example, Uni-Uni-Uni outperforms UniXcoder with an average MRR increase of 9.0\%, and Co-Co-Co achieves an average MRR enhancement of 3.8\% over CoCoSoDa.

\subsubsection{Impact of Weight Selection}

\begin{figure}[!htbp]
	\centering
	\begin{subfigure}{0.45\linewidth}
		\centering

		\includegraphics[width=0.9\linewidth]{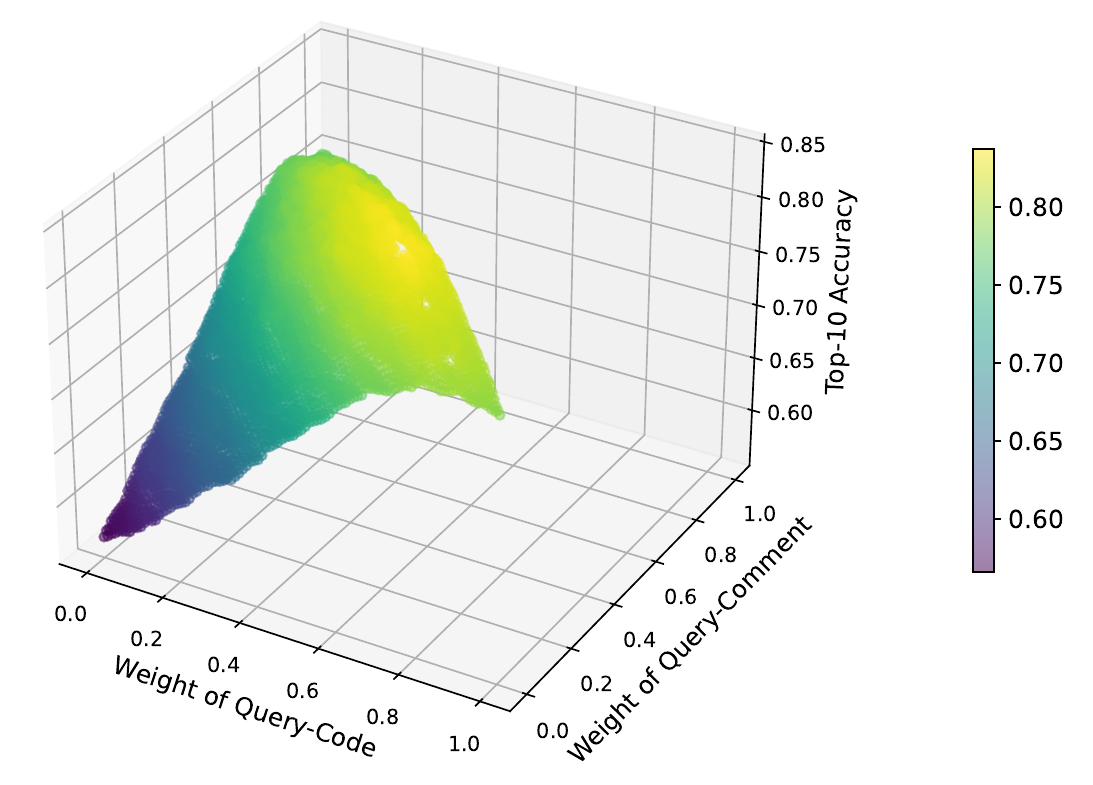}
		\caption{Visualizing the Effects of Weight Adjustments on Top-10 Accuracy}
		\label{fig:3d}
	\end{subfigure}
	\centering
        \hspace{4mm}
	\begin{subfigure}{0.45\linewidth}
		\centering
		\includegraphics[width=0.85\linewidth]{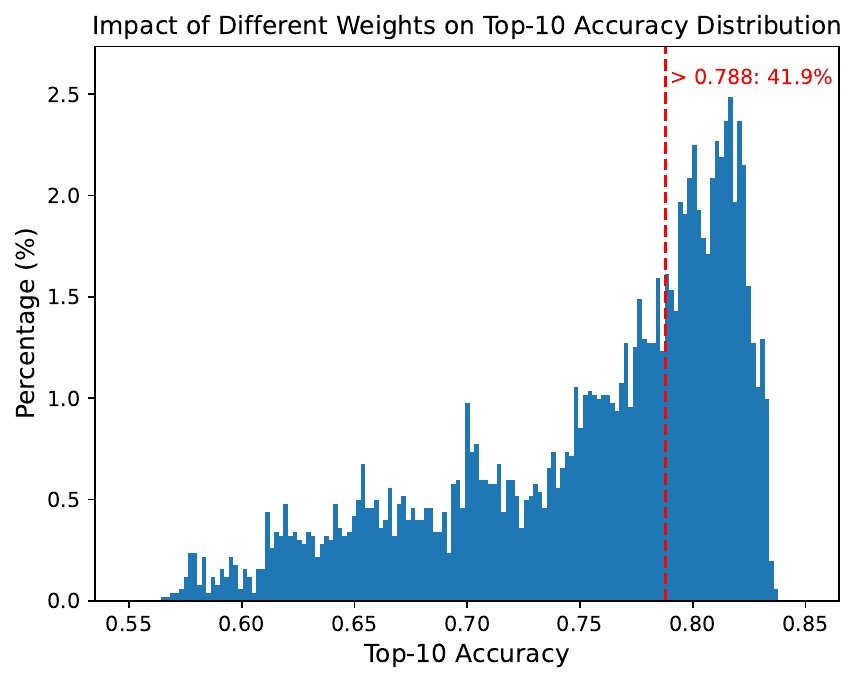}
		\caption{Weight Distribution and Corresponding Top-10 Accuracy}
		\label{fig:bar}
	\end{subfigure}
	\caption{Impact of Weight variations on Top-10 Accuracy: Visualization Results}
	\label{fig:visual}
\end{figure}

\cref{fig:3d} illustrates the variation in top-10 accuracy resulting from different weight adjustments of CodeBridge on the Solidity dataset. 
It can be observed that the variation of top-10 accuracy generally exhibits a semi-peak profile and the peak is not sharp.
This suggests that suitable weights fall within a broad range rather than being critically sensitive to minute changes.
\cref{fig:bar} shows the distribution of the weights that result in various ranges of top-10 accuracy.
Please note that CoCoSoDa achieves a zero-shot performance of 0.788 on top-10 accuracy. 
The result shows that 41.9\% of weight choices can produce outcomes that exceed CoCoSoDa.
This observation is consistent with the data presented in \cref{fig:3d}.
These results suggest that our approach demonstrates robustness in weight selection.

\begin{table}[!htbp]
  \centering
  \belowrulesep=0pt
  \aboverulesep=0pt
    \scriptsize
  \renewcommand\arraystretch{1.2}
  \setlength\tabcolsep{2pt} 
  \caption{Performance on the Solidity Dataset with Different Sampling Datasets}
  \label{tab:sampling-dataset}
  \begin{tabularx}{8.7cm}{C{2cm}|C{1.5cm}|C{1.5cm}|C{1.5cm}|C{1.5cm}}
    \toprule
    Sampling Dataset & MRR & Top-1 & Top-5 & Top-10 \\
    \midrule
    Java/Go/Python& 0.658& 0.571 &0.762& 0.833 \\
    JavaScript& 0.663& 0.579& 0.766& 0.828\\
  \bottomrule
\end{tabularx}
\end{table}

\cref{tab:sampling-dataset} illustrates the performance of our approach with different sample datasets on the Soldiity dataset.
Our approach achieves consistent performance with different sampling datasets.
The optimal weights for the Java, Go, and Python datasets are identical, and our approach achieves even better results with the optimal weight on the JavaScript dataset.
This indicates that our approach is not sensitive to the sampling datasets.

\subsubsection{Different LLMs}
\begin{table}[!htbp]
  \centering
  \belowrulesep=0pt
  \aboverulesep=0pt
  \scriptsize
  \renewcommand\arraystretch{1.2}
  \setlength\tabcolsep{2pt} 
  \caption{Performance on the Solidity Dataset with Different LLMs}
  \label{tab:llms}
  \begin{tabularx}{9.9cm}{C{3.2cm}|C{1.5cm}|C{1.5cm}|C{1.5cm}|C{1.5cm}}
    \toprule
    LLM  & MRR & Top-1 & Top-5 & Top-10 \\
    \midrule
    DeepSeek-Coder-1.3B-Instruct&0.658&0.571&0.762&0.833 \\
    DeepSeek-Coder-6.7B-Instruct&0.664& 0.580& 0.771& 0.824 \\
    CodeLlama-7B-Instruct &0.668 & 0.584 &  0.770 & 0.827 \\
    CodeQwen1.5-7B-Chat& 0.663& 0.576& 0.767& 0.825 \\ 
  \bottomrule
\end{tabularx}
\end{table}
\cref{tab:llms} shows that our approach performs slightly differently with different LLMs, which is attributed to that the zero-shot generation capabilities of different LLMs differ.
It is worth mentioning that our approach outperforms the state-of-the-art baseline CoCoSoDa with different LLMs.
These results indicate that our approach is not sensitive to the used LLMs. 

\rqbox{
\textbf{Answer to RQ6}: Our experiments demonstrate that CodeBridge can perform well across different retrieval models, weight selections, and LLMs.
}\label{finding6}

%% file: tables/rq4_our.tex
\begin{table}[htbp]

  \centering
  \belowrulesep=0pt
  \aboverulesep=0pt
  \renewcommand\arraystretch{1.2}
  \caption{Comparison with Baselines on Cross-domain Setting}
  \scriptsize
  \setlength\tabcolsep{3pt} 
  \label{tab:our-setting}

  \begin{tabularx}{\textwidth}{
  >{\centering\arraybackslash\hsize=1.2\hsize}X
  |*{4}{>{\centering\arraybackslash\hsize=0.5\hsize}X}
  |*{4}{>{\centering\arraybackslash\hsize=0.5\hsize}X}
  |*{4}{>{\centering\arraybackslash\hsize=0.5\hsize}X}}
  \toprule
  \multirow{2}{*}{\textbf{Model}}
& \multicolumn{4}{c|}{\textbf{SQL}} 
& \multicolumn{4}{c|}{\textbf{Solidity}} 
& \multicolumn{4}{c}{\textbf{CoSQA}}   \\

&MRR&Top-1&Top-5&Top-10&MRR&Top-1&Top-5&Top-10&MRR&Top-1&Top-5&Top-10 \\
\midrule
BM25&0.469&0.341&0.614&0.725&0.475&0.391&0.565&0.624&0.183&0.110&0.254&0.312 \\
\midrule
GraphCodeBERT&0.177& 0.120& 0.215& 0.286& 0.196&0.135& 0.249& 0.310&0.089&  0.040& 0.130& 0.190\\
CodeT5+ 110M&0.512&0.385&0.662&0.758&0.424&0.339&0.508&0.579&0.407&0.260&0.586&0.678\\
UniXcoder&0.744&0.632&0.887&0.938&0.544&0.452&0.651&0.701&0.376&0.256&0.512&0.620\\
CoCoSoDa&0.555&0.441&0.690&0.769&0.625&0.541&0.728&0.788&0.482&0.346&0.630&0.724\\
\midrule

\lky{Weighted Mean Pooling} &
\lky{0.115}& \lky{0.077}& \lky{0.138}& \lky{0.180} &
\lky{0.051}& \lky{0.027}& \lky{0.060}& \lky{0.082} &
\lky{0.007}& \lky{0.0}& \lky{0.008}& \lky{0.016}  \\

\lky{Mean Pooling} &
\lky{0.092}& \lky{0.054}& \lky{0.113}& \lky{0.156} &
\lky{0.048}& \lky{0.019}& \lky{0.064}& \lky{0.088} &
\lky{0.006}& \lky{0.002}& \lky{0.004}& \lky{0.008} \\

\lky{EOS Pooling} &
\lky{0.006}& \lky{0.002}& \lky{0.002}& \lky{0.005} &
\lky{0.007}& \lky{0.001}& \lky{0.003}& \lky{0.008} &
\lky{0.004}& \lky{0.002}& \lky{0.002}& \lky{0.008} \\

\lky{Echo Embedding} &
\lky{0.064}& \lky{0.044}& \lky{0.077}& \lky{0.092} &
\lky{0.059}& \lky{0.037}& \lky{0.070}& \lky{0.093} &
\lky{0.046}& \lky{0.024}& \lky{0.058}& \lky{0.088} \\
\midrule
\lky{Query-Code} 
& \lky{0.555} & \lky{0.441} &  \lky{0.690}  & \lky{0.769} 
& \lky{0.625} & \lky{0.541} & \lky{0.728} & \lky{0.788}  
& \lky{0.482} & \lky{0.346} &  \lky{0.630} & \lky{0.724}  \\
\lky{Query-Comment} 
& \lky{0.744} & \lky{0.644} & \lky{0.873} & \lky{0.927}  
& \lky{0.554} &  \lky{0.460} & \lky{0.661} & \lky{0.735}  
& \lky{0.454} & \lky{0.322} & \lky{0.604} & \lky{0.728}  \\
\lky{Code-Code} 
& \lky{0.697} & \lky{0.602} & \lky{0.817} & \lky{0.875}  
& \lky{0.410} &  \lky{0.330} & \lky{0.492} & \lky{0.566}  
& \lky{0.303} &  \lky{0.200} & \lky{0.406} & \lky{0.522}  \\

\midrule
CodeBridge &\textbf{0.811}&\textbf{0.723}&\textbf{0.929}&\textbf{0.958}&\textbf{0.658}&\textbf{0.571}&\textbf{0.762}&\textbf{0.833}&\textbf{0.544}&\textbf{0.424}&\textbf{0.682}&\textbf{0.782}\\
\bottomrule

\end{tabularx}

\end{table}

%% file: tables/rq4_rapid.tex
\begin{table}[htbp]
  \centering
  
    \belowrulesep=0pt
    \aboverulesep=0pt
  \caption{Comparison with baselines on RAPID's setting. The best result is bolded, and the second best is underlined.}
  \scriptsize
  \setlength\tabcolsep{3pt}
  \renewcommand\arraystretch{1.2}
  \label{tab:rapid}
  \begin{tabularx}{\textwidth}{
  >{\centering\arraybackslash\hsize=1.2\hsize}X
  |*{4}{>{\centering\arraybackslash\hsize=0.5\hsize}X}
  |*{4}{>{\centering\arraybackslash\hsize=0.5\hsize}X}
  |*{4}{>{\centering\arraybackslash\hsize=0.5\hsize}X}}

\toprule
\multirow{2}{*}{\textbf{Model}}
& \multicolumn{4}{c|}{\textbf{SQL}} 
& \multicolumn{4}{c|}{\textbf{Solidity}} 
& \multicolumn{4}{c}{\textbf{CoSQA}}   \\

&MRR&Top-1&Top-5&Top-10&MRR&Top-1&Top-5&Top-10&MRR&Top-1&Top-5&Top-10 \\
\midrule
GraphCodeBERT 
&0.175 &0.126 &0.211 &0.257 
&0.292 &0.247 &0.336 &0.369 
&0.500 &0.377 &0.649 &0.740 
\\
+RAPID 
&0.723 &0.628 &0.837 &0.896 
&0.767 &0.710 &0.830 &0.868 
&\textbf{0.899} &\textbf{0.857} &0.956 &0.977
\\
\midrule
UniXcoder
& 0.744 & 0.632 & 0.887 & \underline{0.938} 
& 0.690 &0.623 &0.762 &0.809
& 0.807 &0.737 &0.895 &0.937 
\\
+RAPID
&\underline{0.790} &\underline{0.713} &0.880 &0.923 
&0.779 &0.723 &0.848 &0.876 
&\underline{0.897} &0.846 &\textbf{0.966} &\textbf{0.980} 
\\
\midrule
CoCoSoDa
&0.559 &0.448 &0.691 &0.766 
&0.753 &0.692 &0.822 &0.866 
&0.876 &0.826 &0.945 &0.969 
\\
+RAPID
&0.785 &0.700 &\underline{0.890} &\underline{0.938 }
&\textbf{0.789} &\textbf{0.739} &\textbf{0.853} &\underline{0.878 }
&0.894 &\underline{0.848} &\textbf{0.966} &\textbf{0.980 }
\\
\midrule
CodeBridge
& \textbf{0.811} &\textbf{ 0.723} & \textbf{0.929} & \textbf{0.958} 
& \underline{0.788} & \underline{0.735} & \underline{0.850} & \textbf{0.891 }
& 0.895 & \underline{0.848} & \underline{0.960} & \underline{0.979 }
\\
\bottomrule
\end{tabularx}
\vspace{-3mm}

\end{table}

%% file: sections/discussion.tex
\section{Discussion}\label{sec:discussion}
\subsection{Data Leakage}

LLMs have been trained on open-source data and may have seen the test samples in our experiments. Utilizing LLMs to generate comments and code results in a possibility of data leakage. To mitigate data leakage threats, we additionally collect a dataset containing 1000 Rust samples from GitHub repositories created after February 2023 (the cutoff date for the pre-training data of DeepSeek-Coder). We choose Rust, a language excluded from the embedding model’s pre-training data, to ensure a cross-domain setting. Specifically, we (1) collect 82 non-fork repositories with the largest number of stars; (2) apply the data filtering principles from GraphCodeBERT~\cite{guo2020graphcodebert}, e.g., removing samples with too short or too long queries; and (3) randomly select 1000 samples from the collected 6331 functions.

The results show that CodeBridge achieves an MRR of 0.664, outperforming UniXcoder and CoCoSoDa by 15.7\% and 8.1\%, respectively.
It also outperforms the query-code, query-comment, and code-code matching schemas by 8.1\%, 23.2\%, and 87.6\%, respectively.
The results are consistent with those reported in \cref{sec:performance} across the three datasets.
Therefore, we believe that the threat of data leakage is limited.
Full results can be found in \cite{codebridge}.

\subsection{Computational Efficiency}

To better understand the benefits of CodeBridge and the trade-offs between training and inference overhead, we analyze the computational overhead of CodeBridge compared to the fine-tuning-based approach RAPID (based on UniXcoder). Both RAPID and CodeBridge require offline comment generation, offline codebase processing, online embedding model inference, and online embedding retrieval. However, CodeBridge leverages LLMs to generate code in real time for each query, introducing additional LLM inference overhead, while RAPID fine-tunes UniXcoder with code and generated comments in the target domain, resulting in additional training overhead. As offline comment generation and offline codebase processing are required by both models and are one-time efforts, we focus on their key differences, i.e., training overhead and LLM inference overhead. We also report their online computational overheads to better understand their online efficiency. We use the CoSQA dataset because it has the largest codebase among the three datasets, making it more representative of real-world applications. During training and inference, the maximum lengths are set to 128 for queries and 256 for code. We conduct training using UniXcoder's official training script and utilize vLLM~\cite{kwon2023efficient} for LLM inference. All experiments are performed on one single A800 80GB GPU.  We evaluate two batch sizes, i.e., 64 (the default batch size for UniXcoder training) and 256 (the default maximum batch size for vLLM). To efficiently handle embedding model inference and retrieval, we employ multiprocessing in these two stages.

\input{tables/discussion_efficiency}

\lky{
\cref{tab:efficiency} shows the computational overhead of RAPID and CodeBridge at each stage. The differences in embedding model inference overhead and retrieval overhead between the two models are negligible (less than 1 millisecond per sample). The primary difference lies in the training overhead of RAPID and the inference overhead of LLM. With a batch size of 64, the training of RAPID on one target domain takes 129.0 minutes, and the LLM inference latency of CodeBridge is 36.0 milliseconds per sample, which means the training overhead of RAPID can be used to infer 215,000 samples with the LLM. Increasing the batch size to 256 further raises this capacity to 307,500 samples, but does not significantly speed up embedding model inference and retrieval, as the GPU utilization is already near its maximum. Prior work suggests that a web page response time of less than 2 seconds is considered acceptable for most users~\cite{lohr2012impatient}. The average inference latency of CodeBridge is only about 42 milliseconds per sample. Thus we believe it is acceptable for daily use. With the continuous and rapid advancements in LLM inference acceleration technologies~\cite{zhao2024lookahead,kwon2023efficient}, the inference latency of LLM can be further reduced in the near future.
}

\lky{
While fine-tuning-based methods like RAPID might be cost-effective for a highly popular domain, real-world queries often cover multiple domains (e.g., programming languages)~\cite{guo2024deepseek, sheng2021one}. In such multi-domain scenarios, our one-fit-all framework does not need to maintain and update multiple models for different domains, which is a well-known challenge in machine learning operations and introduces additional costs~\cite{schelter2015challenges}. CodeBridge also benefits less frequently accessed domains. Due to the long-tail distribution~\cite{tiobeTIOBEIndex}, these domains account for the majority and make up a significant share of use cases. For these domains, using CodeBridge is more cost-effective as it eliminates the need for fine-tuning and managing one model for each domain. Thus, we believe CodeBridge provides unique benefits.
}

\subsection{\lky{Code-Comment Pattern}}

\lky{
CodeBridge utilizes both generated code and comments. The code-comment matching, which retrieves comments for the code generated from a query and then returns the associated code, may also be used to improve code search. To investigate this idea, we evaluate code-comment matching using CoCoSoDa, the state-of-the-art embedding model for code search. Experimental results show that code-comment matching performs worst among the four matching schemas on the SQL and CoSQA datasets and second worst on the Solidity dataset, with a relative margin of only 1.5\% over the worst schema. When combining the four matching schemas, grid search reveals that the optimal weights for code-comment matching are consistently zero across all three datasets. Using the sampling method in \cref{sec:sample} to combine the four schemas, the new CodeBridge achieves comparable results to the original CodeBridge on Solidity and CoSQA but suffers a 2.1\% MRR drop on SQL, likely due to the noise in the generated code and comments. Therefore, this matching schema does not bring significant benefits considering its performance and additional overhead. The full results are available at \cite{codebridge}.
}

%% file: tables/discussion_efficiency.tex
\begin{table}[!htbp]
    \vspace{-3mm}
  \centering
  \belowrulesep=0pt
  \aboverulesep=0pt
  \color{\tablecolor}
  
  \renewcommand\arraystretch{1.2}
    \scriptsize
  \caption{\lky{Training and inference overhead of different models}}
  \setlength\tabcolsep{2pt} 
  \label{tab:efficiency}
  \begin{tabularx}{\textwidth}{
  >{\centering\arraybackslash\hsize=0.15\hsize}X
  |>{\centering\arraybackslash\hsize=0.1\hsize}X|
  *{4}{>{\centering\arraybackslash\hsize=0.1875\hsize}X}}
  \toprule
   Model & Batch Size & Training \textbf{(min}) & LLM Inference (\textbf{ms/sample}) & Embedding Model Inference (\textbf{ms/sample}) & Retrieval (\textbf{ms/sample}) \\
  \midrule
    RAPID (UniXcoder) &  \multirow{2}{*}{64} & 129.0 & --- & 2.7 & 2.7 \\
    CodeBridge & &  --- & 36.0 & 3.2 & 2.8 \\
    \midrule
     RAPID (UniXcoder) &  \multirow{2}{*}{256} & 123.0 & --- & 2.6 & 2.7 \\
    CodeBridge & &  --- & 24.0 & 3.0 & 2.7 \\
\bottomrule

\end{tabularx}
    \vspace{-3mm}

\end{table}

%% file: sections/threats.tex
\section{Threats to Validity}

The generalizability of our experimental results is a threat to the validity of this study.
Due to practical constraints, we are unable to assess the cross-domain performance of our approach across all languages. 
However, we test three languages, including two less common ones (SQL and Solidity), which are not part of the languages in the training data for the PLMs we use.
Experimental results show that our approach exhibits consistent performance across the three datasets, showcasing its language-agnostic nature and demonstrating its applicability across various domains.

%% file: sections/conclusion.tex
\section{Conclusion}
In this paper, we conduct an empirical study to reveal the effectiveness of query-comment matching and code-code matching in zero-shot cross-domain code search for the first time.
The empirical results reveal the high complementarity among three matching schemas, i.e., query-code, query-comment, and code-code matchings. 
Based on the empirical findings, we propose CodeBridge, a zero-shot cross-domain code search approach without fine-tuning.
It first leverages LLMs to generate comments and code to expand the query-code pairs and then integrates the three matching schemas.
We also propose a sampling-based fusion approach to combine the three similarity scores to rank the code.
Experimental results show that CodeBridge outperforms the state-of-the-art PLM-based code search approaches and yields results that are better than or comparable to those of the zero-shot cross-domain code search technique RAPID, which requires fine-tuning.
In the future, we plan to enhance the generation quality for retrieval and ways to improve efficiency.

%% file: sections/data_avaliability.tex
\section{Data Availability}
Our replication package, including datasets and source code, is available at \cite{codebridge}.